\documentclass[letter]{aa} % for the letters 
\usepackage{caption}
\usepackage{color}
\usepackage{subcaption}
\usepackage{multirow}% http://ctan.org/pkg/multirow
\usepackage{mathrsfs}

\usepackage{graphicx}
\usepackage{txfonts}
\def\mso{\,M_\odot}

\def\Teff{T_{\rm eff}}

\def\simgr{\,\hbox{\hbox{$ > $}\kern -0.8em \lower 1.0ex\hbox{$\sim$}}\,}
\def\simle{\,\hbox{\hbox{$ < $}\kern -0.8em \lower 1.0ex\hbox{$\sim$}}\,}

\begin{document} 

   \title{The spectroscopic Hertzsprung--Russell diagram of Galactic massive stars}

   \author{N.~Castro\inst{1},
   		   L.~Fossati\inst{1},
   	       N.~Langer\inst{1},
           S.~Simón-Díaz\inst{2,3}, 			 		   	 
   	   	   F.~R.~N.~Schneider\inst{1} and
		   R.~G.~Izzard\inst{1}
          }
     \institute{Argelander-Institut für Astronomie der Universität Bonn, Auf dem Hügel 71, 53121, Bonn, Germany\\
              \email{norberto@astro.uni-bonn.de}
         \and
            Instituto de Astrofísica de Canarias, 38200, La Laguna, Tenerife, Spain 
       \and       
        Universidad de La Laguna, 38205, La Laguna, Tenerife, Spain
             }
   \date{Received --; accepted --}
\titlerunning{The Galactic sHRD.}
\authorrunning{N. Castro et al.}
 
  \abstract
   {The distribution of stars in the Hertzsprung--Russell diagram narrates their evolutionary history and directly assesses their properties. Placing stars in this diagram however requires the knowledge of their distances and interstellar extinctions, which are often poorly known for Galactic stars. The spectroscopic  Hertzsprung--Russell diagram (sHRD) tells similar evolutionary tales,  but is independent of distance and extinction measurements. Based on spectroscopically derived effective temperatures and gravities of almost 600 stars, we derive for the first time the observational distribution of Galactic massive stars in the sHRD. While biases and statistical limitations in the data prevent detailed quantitative conclusions at this time, we see several clear qualitative trends. By comparing the observational sHRD with different state-of-the-art stellar evolutionary predictions, we conclude that convective core overshooting may be mass-dependent and, at high mass ($\simgr 15\mso$), stronger than previously thought. Furthermore, we find evidence for an empirical upper limit in the sHRD for stars with $T_{\rm{eff}}$ between 10\,000 and 32\,000\,K  and, a strikingly large number of objects below this line. This over-density may be due to inflation expanding envelopes in massive main-sequence stars near the Eddington limit. }


   \maketitle
%
%________________________________________________________________

\section{Introduction}

The Hertzsprung--Russell diagram \citep[HRD,][]{He1905,1919PNAS....5..391R} is probably the most important tool for analysing stellar evolution \citep{1964Cent....9..219N}. The main-sequence is the most prominent feature in the HRD, identifying the core hydrogen burning stars and allowing us to distinguish the later evolutionary stages \citep[e.g.][]{2003ARA&A..41...15M,2010ARA&A..48..581S}. However, the stellar distance and reddening are required to derive the stellar luminosity, which may render uncertain the position of, e.g., Galactic stars in the HRD. \cite{2014A&A...564A..52L} proposed an alternative tool for analysing physical properties of observed stars and testing stellar evolution models, the spectroscopic  Hertzsprung--Russell diagram (sHRD). The sHRD is obtained from the HRD by replacing the luminosity ($L$) to the quantity  ${\mathscr L} := T_{\rm eff}^4/g$, which is the inverse of the flux--weighted gravity introduced by \cite{2003ApJ...582L..83K}. The value of ${\mathscr L}$  can be calculated from stellar atmosphere analyses without prior knowledge of the distance or the extinction. In contrast to the classical $T_{\rm{eff}}-{\rm log}\,g$ diagram (Kiel diagram), the sHRD sorts stars according to their proximity to the Eddington limit, because ${\mathscr L}$ is proportional to the Eddington factor $\Gamma_{\rm e} = L/L_{\rm Edd}$,
\begin{equation}
{\mathscr L} = {1\over 4\pi\sigma G}{L\over M} = {c\over \sigma \kappa_{\rm e}} \Gamma_{\rm e} ,
\end{equation}
where $\sigma$ is the Stefan-Boltzmann constant, $\kappa_{\rm e}$ is the electron scattering opacity in the stellar envelope, and the other symbols have their usual meanings.

Except for their final stages, the evolution of low and intermediate mass stars, including our Sun,
 is well understood \citep[e.g.][]{1985ApJS...58..711V,2000A&AS..141..371G}, as their initial mass largely settles their
evolution and fate \citep{1967ARA&A...5..571I}.
In massive ($ \simgr 8\mso$) stars, however, even the core hydrogen burning phase is
not yet well understood and the knowledge about core helium burning evolution stages is even more uncertain
\citep{2012ARA&A..50..107L}. The higher the considered  masses the less we know about how
the stars evolve. The main reason for this is that, besides their initial mass,
the evolution of massive stars depends on further initial parameters:
 spin  \citep{2000A&A...361..101M,2000ApJ...528..368H} 
and duplicity  \citep{2012Sci...337..444S}.
Furthermore, uncertain mass loss rates \citep{2014arXiv1402.1237S} and internal mixing processes
\citep{1981A&A...102...25B,1991A&A...252..669L,1992A&AS...96..269S,2006A&A...460..199Y} are more important the more massive the stars.
For these reasons, our knowledge of the evolution of massive stars is rudimentary and observational constraints are urgently needed.

To exploit the capacities of the sHRD for massive stars, 
we have mined the available quantitative spectroscopic studies 
in the literature, and augmented this with our own analyses of luminous OB stars, to collect
accurate effective temperatures and gravities of a large number of Galactic massive stars.
This allows us to establish  for the 
 first time the distribution of these objects
in the sHRD. After discussing our stellar sample and its biases in Sect.\,2,
we show an empirical sHRD in Sect.\,3, where we also compare it with predictions from
recent stellar evolution models. In Sect.\,4, we discuss our results and present conclusions.

 	\begin{figure*}
   		\resizebox{\hsize}{!}{\includegraphics[angle=90,width=\textwidth]{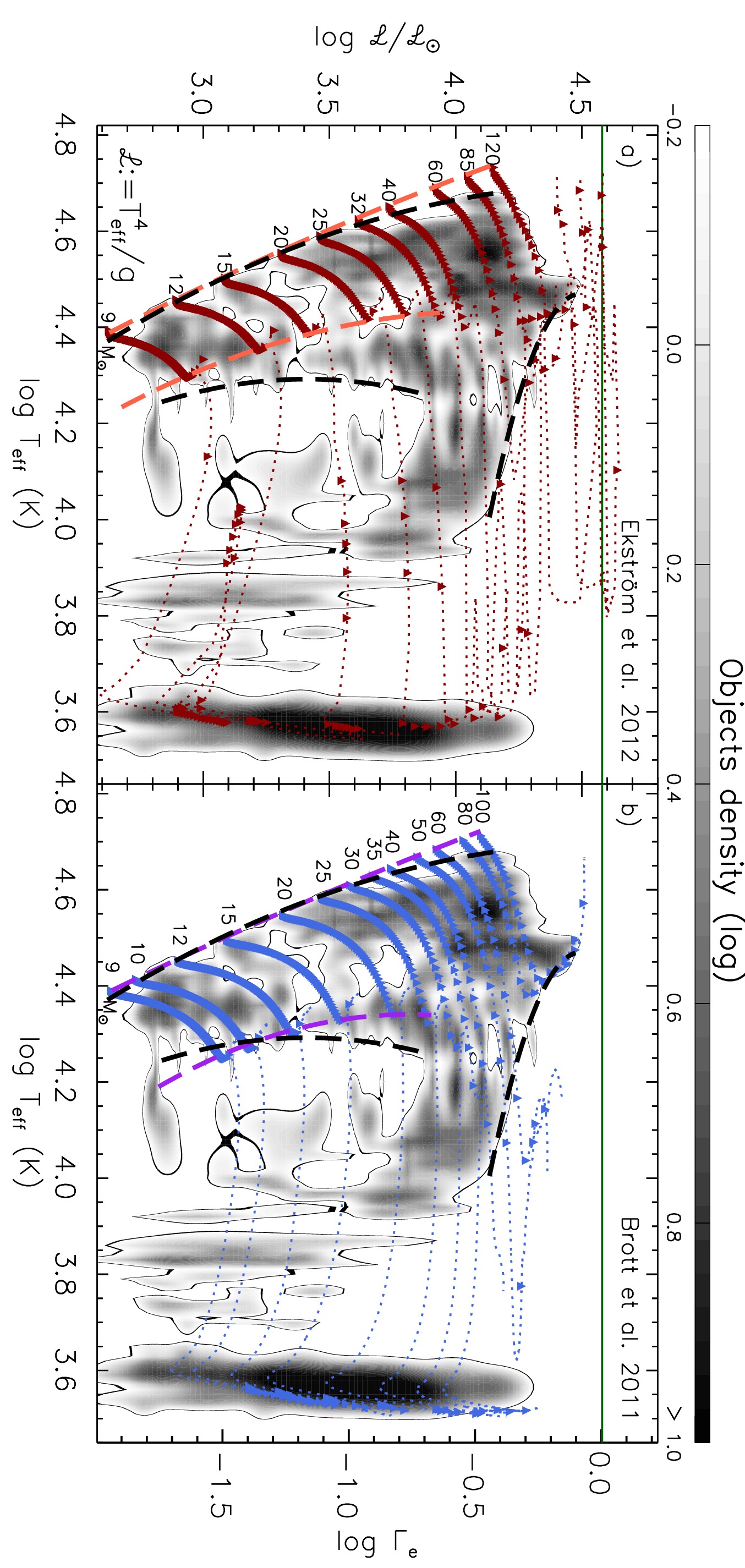}}
   		\caption{Grey scale representation of the probability density distribution of the location of 575 Galactic 
stars in the sHRD. Three empirical borderlines between densely populated regions and empty regions are drawn as black dashed lines (cf.  Table~\ref{TAB:boun}).
 The electron scattering Eddington factor ($\Gamma_{\rm e}$) is given on the right ordinate axis. 
The $\Gamma_{\rm e}$ limit for hydrogen-rich composition at $log\,{\mathscr L}/{\mathscr L}_\odot=4.6$ is represented 
by a green horizontal line. Overlayed are stellar evolution tracks for non-rotating stars with solar composition: a)  \citet{2012A&A...537A.146E} and, b) \citet{2011A&A...530A.115B}. The ZAMS and TAMS positions of the
models are connected through orange and purple dashed lines. Red and blue triangles are
placed on the tracks separated by $0.1\,$Myr. }

\label{Fig:SHRD_blue}
\end{figure*}

\section{Stellar sample}
Our sample comprises 575 stars, 439 of which have $T_{\rm{eff}}\ge10\,000$\,K and 
log\,${\mathscr L}/{\mathscr L}_\odot>2.75$ (corresponding to masses above $\sim9\mso$). The analyses of 255 stars belong to the IACOB
spectroscopic survey of Northern Galactic OB-stars project \citep{2011BSRSL..80..514S,2014arXiv1409.2416S}, using the techniques described in \cite{2011JPhCS.328a2021S} and \cite{castro2011}.
Parameters for another 250 stars were extracted 
 from  the literature 
 \citep[][]
{1999A&A...349..553M,2001A&A...379..936P,2004A&A...413..693M,2014A&A...562A..37M,2004A&A...415..349R,2005ApJ...628..973L,2005A&A...441..735M,2012A&A...538A..39M,   2005A&A...441..711M,2006A&A...446..279C,2007CoAst.150..183B,2007A&A...463.1093L,
2010A&A...515A..74L,2008A&A...481..777S,2009A&A...496..841H,2010A&A...510A..22S,
2012A&A...543A..80F,2013A&A...550A..26N,2014ApJ...781...88A}. We complemented the sample with objects from the catalogue gathered by 
\cite{2010A&A...515A.111S} with known metallicity above $\mathrm{[Fe/H]}=-0.5$. 

\subsection{Completeness and biases}
The collated data comprise a heterogeneous sample in terms of instruments used for the observations, tools, and methodologies 
adopted for the spectral analysis. The most prominent characteristic of our sample is that it is 
 dominated by bright stars. For instance, \cite{2010A&A...515A.111S} pointed out that 90\% of the stars listed in their catalogue are brighter than $V=9.75$\,mag. The analysis of OB stars is also dominated by bright stars in the solar neighbourhood.

One important bias that affects the sample is the authors' interest: different authors focus on a particular kind of star to tackle a particular astrophysical topic. This leads to a spurious  overpopulation of stars in certain regions (e.g. pulsational instability domains, abundances in early B-type stars or the interest in the O-type and early B-supergiant regimes) compared to others.

This melting pot approach  introduces biases that cannot be quantified. Nevertheless, the sample has its strength in the large number 
of stars that it contains. Although it is far from giving a statistical view of the Milky Way, 
it is large enough to minimise some of the biases and to enable the determination of well populated regions and 
its borders across the upper sHRD, with the aim of constraining stellar evolution models. 

\section{The sHRD of massive stars}
We determine the position of each star in the sHRD on the basis of its atmospheric parameters. 
We derive the probability density function by adding the Gaussian probability distributions of the stellar 
parameters, which are derived from their reported uncertainties. 
In cases where the uncertainties are not provided, we adopted typical error bars based on the studies 
compiled in this work. For each star, the probability distribution was calculated adopting a 
Gaussian distribution in the log$\,{\mathscr L}/{\mathscr L}_\odot-\,$log$\,T_{\rm{eff}}$ parameter space. 
All distributions are summed to give the density map shown in Fig.~\ref{Fig:SHRD_blue}. In addition, Fig.~\ref{Fig:SHRD_blue} shows stellar evolution tracks for non-rotating, solar-metallicity single stars published by \cite{2011A&A...530A.115B} and \cite{2012A&A...537A.146E}, and  polynomial 
fits to the theoretical zero age main-sequence (ZAMS) and the terminal age main-sequence (TAMS) lines.

While the probability density map is somewhat patchy, it allows us to identify several borderlines that 
separate regions of high density from those where stars are almost absent.
Three such lines are drawn in Fig.~\ref{Fig:SHRD_blue}. One nearly vertical borderline 
identifies the location to the left of which no stars are found (note that Wolf-Rayet stars were
not included in our sample), which could therefore relate to the ZAMS.
A second nearly vertical borderline, located at log\,$\Teff\,($K$)\simeq 4.3$, could be
the TAMS. Finally, there is a close to horizontal borderline
near  log\,${\mathscr L}/ {\mathscr L}_\odot \simeq 4.3$, which corresponds to an electron scattering Eddington factor of $\Gamma_{\rm e}\simeq 0.5$.  
These shown borderlines are quadratic fits (Table~\ref{TAB:boun}).
Figure~\ref{Fig:SHRD_blue} shows an underdensity of stars across the main-sequence in the $15-25\mso$ range.
We ascribe this anomaly to an observational bias, but additional observations in this region are necessary.

\begin{table}
  \caption{Polynomial coefficients of the observed and theoretical boundaries marked in Fig. \ref{Fig:SHRD_blue}.  }
  \label{TAB:boun}
  \begin{tabular}{lrrrrr}
  \hline
Empirical &   $Y_{min}$  &  $Y_{max}$   &        $a$ &       $b$ &       $c$ \\

\hline
ZAMS &     2.62 &  4.19 &      2.973 &     0.743 &    -0.080 \\
TAMS &    2.84 &   3.88 &      2.618 &     0.982 &    -0.144 \\
UPPER &     4.13 &  4.48 &    -67.323 &    31.938 &    -3.552 \\
\hline
\multicolumn{4}{l}{Ekstr\"om et al. 2012 (non-rotating)}      &     &     \\
\hline
ZAMS &          2.63 &  4.14 &      3.513 &     0.399 &    -0.025 \\
TAMS &          2.68 &  3.94 &      2.816 &     0.784 &    -0.095 \\
\hline
\multicolumn{4}{l}{Brott et al. 2011 (non-rotating)}       &     &     \\
\hline
ZAMS &  2.64 &  4.10 &      3.586 &     0.355 &    -0.019 \\
TAMS &  2.82 &  3.91 &      2.103 &     1.174 &    -0.154 \\

\hline

\end{tabular}
\tablefoot{$T_{\rm{eff}}\,(K)=c\,Y^2+b\,Y+a$, where $Y=\,$log$\,{\mathscr L}/{\mathscr L}_\odot$. $Y_{min}$ and  $Y_{max}$ set the boundaries of quadratic fits. }
\end{table}

\subsection{The mass range $8-30\mso$}
\label{730M}
The $8-30\mso$ range covers stars with spectral types in the main-sequence between B2 and O7. 
Figure~\ref{Fig:SHRD_blue} shows a good match between the observed ZAMS and both 
sets of evolutionary tracks. Note that, although the initial solar composition adopted by \cite{2011A&A...530A.115B} and \cite{2012A&A...537A.146E} is slightly different, it is the iron abundance with which the opacities are interpolated in the \cite{2011A&A...530A.115B} models.
Since the adopted iron abundances are the same, the positions of the two ZAMS lines are nearly identical.

The stars are mainly clustered on the main-sequence, where they are expected to remain for most of their lives. 
Our density map shows a rather tight boundary on the cool side of the main-sequence,
which could be interpreted as the empirical position of the TAMS.
In the following, we will use this interpretation as our working hypothesis.

The width of the main-sequence band predicted by \cite{2012A&A...537A.146E} 
fits the observations well around $8\mso$, but the observed main-sequence becomes  
wider with increasing mass up to $30\mso$. This is not predicted by \cite{2012A&A...537A.146E}. 
On the other hand, \cite{2011A&A...530A.115B} predict a main-sequence that is too wide around $8\mso$, 
 fits well around $\sim15\mso$ and then worse at higher mass.

The difference between the evolutionary models might be caused by the different calibration of the
 convective core overshooting parameter  $d_{\mathrm over}$. 
The value of  $d_{\mathrm over}/H_{\rm P}=0.10$ adopted by \cite{2012A&A...537A.146E} was calibrated using tracks for
rotating stars in an intermediate mass range from 1.35 to $8\mso$.
  On the other hand, $d_{\mathrm over}/H_{\rm P}=0.335$ 
was used by \cite{2011A&A...530A.115B}, as derived by comparing stars around $16\mso$. 
Figure~\ref{Fig:SHRD_blue} shows that even more overshooting may be necessary at higher mass. In fact, Fig.~\ref{Fig:SHRD_blue} argues for a mass-dependent overshooting.

Figure~\ref{Fig:SHRD_blue} shows the tracks of non-rotating models, but using tracks for rotating stars 
does not solve the discrepancy (Sect.~\ref{sec:rotation}). It is possible that the 
mass-dependence of the discrepancy between the observed and theoretical TAMS lines is due to yet unconsidered physics, 
possibly connected to rotation, which is not included in either set of tracks, 
rather than mass-dependent overshooting.

On the red side of the TAMS, Fig.~\ref{Fig:SHRD_blue} shows a gap between the TAMS and 
 blue supergiants of log\,$\Teff\,($K$)\simle 4.15$. This post main-sequence gap, known as Hertzsprung gap, 
is in agreement with stellar tracks, 
which predict that stars rush through this gap in a fast evolution. The stars found at log\,$\Teff\,($K$)\simle 4.15$
can be interpreted as core helium burning objects as predicted by models of \cite{2012A&A...537A.146E}.
Note that the models of \cite{2011A&A...530A.115B} were stopped before the tracks reached this evolutionary phase.
The number of objects detected in this region of our empirical sHRD could be affected 
by interest bias because of the presence of pulsations \citep[e.g.][]{2007CoAst.150..183B}. 

The region covered by the red supergiant stars is reproduced by both sets of models. 
While the empirical temperature scale of the red supergiants may be hotter than previously thought,
it is still uncertain \citep{2005ApJ...628..973L,2013ApJ...767....3D}. In the stellar models, it is mainly the mixing length parameter that controls the temperature of the
Hayashi line \citep{1990sse..book.....K}.  

\subsection{Stars above $30\mso$}
\label{30M}
The high mass part of both theoretical ZAMS lines has a clear
offset from the empirical ZAMS, which increases with mass. 
While we cannot exclude bias effects, as the number of observed very massive stars
near the ZAMS is small, the complete absence of stars near the upper end of the theoretical ZAMS
lines might also be related to the fact that such young massive stars are likely still embedded 
in their birth clouds \citep{1986ARA&A..24...49Y}. 

In the log$\,T_{\rm{eff}}\,($K$)\simeq 4.0-4.5$ range, the observations set an upper boundary in the sHRD 
(Table~\ref{TAB:boun}), reminiscent of the Humphreys-Davidson 
limit in the HRD \citep{1979ApJ...232..409H}. We recall here that we did not include 
luminous blue variable stars or stars with emission line dominated spectra in our sample.
Figure~\ref{Fig:SHRD_blue} shows that the tracks of \cite{2011A&A...530A.115B} fit the upper boundary                   
of the main-sequence massive stars (log$\,T_{\rm{eff}}\,($K$)\simeq 4.3-4.5$), which was interpreted 
as an effect of the true Eddington limit by \cite{2012ARA&A..50..107L} (rather than the electron scattering Eddington limit). 

The green line in Fig.~\ref{Fig:SHRD_blue} corresponds to the electron scattering Eddington limit for solar hydrogen abundance. Stars with a helium-rich surface may be located above the line, because the Eddington limit for pure helium composition lies 0.23 dex above the green line. The \cite{2012A&A...537A.146E} models extend to this region because they include the Wolf-Rayet phases.

Figure~\ref{Fig:SHRD_blue} shows that the red supergiant branch stretches up to larger Eddington factors than the blue supergiant region. This may occur because in red supergiants a significant fraction of stellar luminosity is transported by convection rather than by radiation even out to the photospheric layers. They can thus bare larger Eddington factors even if the opacities are larger than in blue supergiant envelopes.

Below the line related to the Eddington limit, stars are distributed continuously
from the main-sequence to surface temperatures just below 10\,000\,K. In the mass range considered here,
a TAMS line cannot be identified from the data. This could imply that the main-sequence band
above $30\mso$ extends to temperatures below 10\,000\,K. This is predicted by the tracks of \cite{2011A&A...530A.115B},
even though at slightly larger mass-to-luminosity ratios, and is found to be a consequence of the so called
envelope inflation that occurs as a result of the proximity to the Eddington limit \citep[][]{kohler2014}. 
Alternatively, this population of blue super- and hypergiants might consist of core helium burning stars 
\citep[see][]{1995A&A...295..685L}. 
Whereas their large number may be in conflict with this interpretation, we cannot rule out that some interest bias 
 affects the observed distribution, because of the 
work done to model A- and B-supergiants \citep[e.g.][]{2012A&A...543A..80F}.

\subsection{Rotation}
\label{sec:rotation}
Even during the main-sequence phase, the evolution of massive stars may be significantly affected by rotation 
\citep{2000A&A...361..101M,2000ApJ...528..368H}. 
As our sample is composed of stars with a large variety of rotational velocities, 
it could be important to consider stellar tracks for rotating stars also. 
We therefore show in Fig.~\ref{Fig:SHRD_blue_rot_tofo} a comparison of tracks for non-rotating and rapidly
rotating tracks from the same model grids in Fig.~\ref{Fig:SHRD_blue}. 
We do not show models with even faster rotation, because these stars are rare \citep{1977ApJ...213..438C,2014A&A...562A.135S}.

While the effects of rotation on the position of the ZAMS are negligible within the scope of our paper,
the TAMS positions of the rapidly rotating models are indeed somewhat shifted with respect to the 
non-rotating models. However, when compared to the observational data, the conclusions drawn in the previous
subsection remain unaltered. In the \cite{2012A&A...537A.146E} models, the TAMS line for stars
below $30\mso$ changes very little, and for more massive stars the main-sequence band becomes 
even narrower with rotation. In the \cite{2011A&A...530A.115B} models, the main-sequence band widens for
stars below $15\mso$, which increases the discrepancy with observations. Overall, considering rapid rotation
appears not to bring the models into any better agreement with the observations.
\section{Conclusions}

We present the first observational spectroscopic Hertzsprung--Russell diagram (sHRD) 
for Galactic massive stars ($\simgr 8\mso$), 
based on the spectroscopic analysis and atmospheric modelling of a sample of 575 stars.
We produce a probability density map of the positions of stars in the sHRD.
This map shows several clear borderlines that provide stringent constraints on massive star evolution models. Most notably,
these lines may correspond to the ZAMS and TAMS borders of the main-sequence band.
Additionally we find a sharp upper limit of observed stars in the sHRD near but somewhat below
the electron scattering Eddington limit.

Except for the most massive stars, for which early hydrogen burning evolution may be hidden by their birth cocoons,
the models represent the ZAMS position in the sHRD well. However, neither the models
of \cite{2011A&A...530A.115B} nor those by \cite{2012A&A...537A.146E} can reproduce what we tentatively identify as the
TAMS in the sHRD. One possible interpretation is that the convective core
overshooting parameter in massive stars increases with mass.

A further striking feature in our probability map is a well populated area just below
the sharp upper mass-to-luminosity ratio limit in the sHRD that extends from the main-sequence to beneath
10\,000\,K. While none of the studied stellar evolution models can fully reproduce these
stars, they may be interpreted as stars with inflated envelopes because of their proximity
to the Eddington limit in the frame of the \cite{2011A&A...530A.115B} models. 

We consider this work to be a  first step towards providing essential constraints for the
evolution of massive and very massive stars.
Indeed, we are far from a complete view of the Milky Way stellar content, and our sample
contains voids that require additional data. Several on-going surveys 
of OB-type stars will increase the number of objects in the sHRD in the next few years and allow us a
more detailed and robust comparison with stellar evolution models.

\begin{acknowledgements}
The authors thank the referee for useful comments and helpful suggestions that improved this manuscript. LF and RGI acknowledge financial support from the Alexander von Humboldt Foundation. SS-D thanks funding from the Spanish Government Ministerio de Economia y Competitividad (MINECO) through grants AYA2010-21697-C05-04, AYA2012-39364-C02-01 and Severo Ochoa SEV-2011-0187, and the Canary Islands Government under grant PID2010119. FRNS acknowledges the fellowship awarded by the Bonn--Cologne Graduate School of Physics and Astronomy.
\end{acknowledgements}

%-------------------------------------------------------------------

%############################################################
%	BIBLIOGRAPHY
%############################################################

\bibliographystyle{aa}

\bibliography{NC_AA_2014_25028}

%############################################################
%       Appendix
%############################################################

\begin{appendix}
\section{sHRD with rotating models}

\begin{figure*}
\resizebox{\hsize}{!}{\includegraphics[angle=90,width=\textwidth]{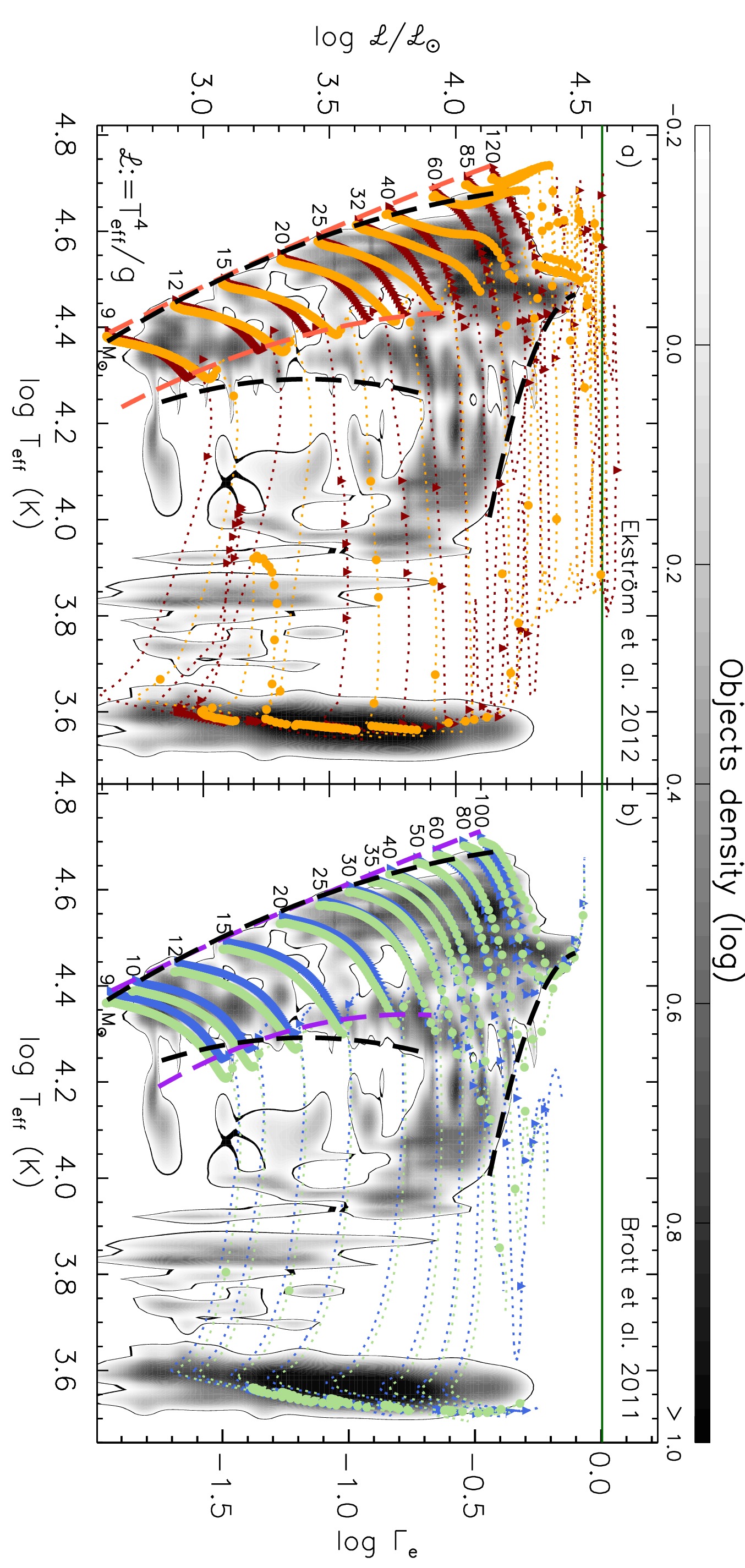}}
\caption{As in Fig. \ref{Fig:SHRD_blue}, but including rotating stellar evolution tracks:  
a) models of \cite{2012A&A...537A.146E} with an initial rotational velocity of 40\% of critical rotation (orange 
dotted lines and dots) and, b) models of \cite{2011A&A...530A.115B} with an initial rotational velocity 
of 300\,km\,s$^{-1}$ (green dotted lines and dots). }
\label{Fig:SHRD_blue_rot_tofo}
\end{figure*}
\end{appendix}

\end{document}